\documentclass[11pt,a4paper]{article}
\usepackage{cfm2009}

\usepackage{psfrag}
\usepackage{epsfig}
\usepackage{latexsym}

\renewcommand{\vec}[1]{\mbox{\boldmath $#1$}}

\newcommand{\pd}[1]{\partial_{#1}}
\renewcommand{\vec}[1]{\mbox{\boldmath $#1$}}

\begin{document}
\cfmtitre{Coherent structures in fully-developed pipe turbulence}
\cfmauteur{A.\ P.\ Willis,\,\, Y.\ Hwang,\,\, C.\ Cossu}
\cfmadresse{
   Laboratoire d'Hydrodynamique (LadHyX),  \'Ecole Polytechnique, 91128 PALAISEAU (France)
}

\cfmresume{
Un profil moyen turbulent est prescrit dans une conduite cylindrique, en ad\'equation avec les observations exp\'erimentales. Nous consid\'erons ensuite la nature des perturbations \`a cet \'ecoulement synth\'etique. Le calcul des croissances optimales pr\'edit deux types de structures,  associ\'ees respectivement aux structures de proche-paroi et de grande \'echelle.  Un excellent accord quantitatif est trouv\'e avec les r\'esultats exp\'erimentaux quant \`a la longueur d'onde transversale. La r\'eponse harmonique est \'egalement \'etudi\'ee, et la croissance lin\'eaire observ\'ee compar\'ee \'a des simulations num\'eriques directes de turbulence forc\'ee. Malgr\'e de l'hypoth\`ese simple de type `Eddy viscosity', cette approche lin\'eaire pr\'edit efficacement la croissance spectaculaire des modes de grande \'echelle au coeur de l'\'ecoulement. 
}

\cfmabstract{
A turbulent mean profile for pipe flow is prescribed which closely matches experimental observations.  The nature of perturbations superimposed upon this profile is then considered.  
Optimal growth calculations predict two distinct classes of structures, clearly associated with near-wall and large-scale structures.  Quantitative correspondence of the spanwise wavelength 
of wall-structures with experimental observations is very good.  The response to harmonic
forcing is also considered, and the linear growth tested with direct numerical simulation of forced turbulence.  Despite the very simple eddy viscosity assumption, 
this linear approach predicts well the surprisingly large growth of 
outer-scale modes in the bulk flow.
}

\cfmmotscles{Turbulence, Optimal growth, Pipe flow}
\section{Introduction}

The discovery of nonlinear solutions to the Navier--Stokes equations for
Pipe and Couette geometries has led to significant progress of late.
For pipe flow, these travelling wave solutions were discovered by 
adding a forcing to the Navier--Stokes equations to induce axially independent 
rolls.  As the forcing is increased a three-dimensional wave instability arises.  
Three-dimensional states could then be traced back, using continuation methods, to the 
original equations upon reduction of the forcing
\cite{faisst03,wedin04}.
Whilst successful at finding the first nonlinear states, the nature of this
method leads to solutions that are only just self-sustained.  As a consequence,
the majority of known solutions (of the sinuous variety) were shown to be
characteristic of the boundary between laminar and turbulent states.  
Initial states either side of this boundary develop into either turbulence or 
relaminarise \cite{kerswell07,itano01,wang07,duguet08}.  
Only later were higher branches found (being also highly symmetric) 
with higher friction factors typical of turbulence
\cite{pringle09}.
The highly organised structure of these states, however, does not appear 
to be typical of turbulence, nor do they display the strongly 
decelerated core of the mean flow.  It is perhaps more intuitive,
therefore, that one should begin with perturbations to this mean flow.

\section{Method}

Here we consider perturbations to a close approximation to the mean
flow profile for pipe flow.  The equations are normalised by the 
bulk flow speed $U_b$ and the radius of the pipe $R$.  Reynolds numbers
are defined $Re=2\, U_bR/\nu$ and $Re_\tau=u_\tau R/\nu$, where the wall
velocity $u_\tau=(\nu\,\pd{r}U(r))^{\frac{1}{2}}|_{r=R}$ is derived
from the mean stress at the boundary.
The mean profile $U(r)$ is inverted from the 
$\theta$,$z$-averaged Navier--Stokes equations
\begin{equation}
   0 = -\pd{z}P + \frac{1}{Re}
   \left(
      \frac{1}{r} + \pd{r}
   \right)
   (\nu_T \pd{r} U ),
   \qquad
   B = -\pd{z}P ,
\end{equation}
where the normalised effective viscosity is $\nu_T(y)=1+E(y)$,
and $y=1-r$ is the distance from the pipe wall.
The radially dependent eddy viscosity is prescribed
(\cite{reynolds67}, after \cite{cess58}), 
\begin{equation}
   E(y) = \frac{1}{2}
   \left\{
      1 + \frac{\kappa^2 \hat{R}^2 \hat{B}}{9}
      [2y-y^2]^2
      (3-4y+2y^2)^2
      \left[
         1 - \exp
         \left(
            \frac{-y \hat{R} \sqrt{\hat{B}}}{A^+}
         \right)
      \right]
   \right\}^{\frac{1}{2}} - \frac{1}{2} \, .
   \nonumber
\end{equation}
We have updated the fitting parameters $A^+=27$ and $\kappa=0.42$ to be
compatible with the observations of
\cite{mckeon05} and different scalings lead to the adjusted
parameters $\hat{R}=Re\,/\,2$,\, $\hat{B}=2\,B$.

Perturbations, $\vec{u}$, to the mean profile are first considered in the
linear framework.  The Navier--Stokes equations with radially-dependent 
viscosity are linearised about $U(y)$; note that the perturbations about 
$U$ are also subject to the elevated effective viscosity.
Rather than the usual progression to Orr--Sommerfeld--Squire form, 
eigenvalues and eigenvectors are found directly from the linearised 
primitive variable system with explicit solenoidal condition.
Given the eigenfunctions and eigenvalues, 
using standard methods \cite{schmid94}
one may calculate the optimal growth,
\begin{equation}
   G(\alpha,m;t) = \max_{\vec{u}_0} \, \frac{||\vec{u}(t)||}{||\vec{u}_0||}
 \quad \mbox{ and } \quad
   G_{\mathrm{max}} = \max_t {G(t)} .
\end{equation}
The wavelengths of the perturbations are $\lambda_z = 2\pi/\alpha$ (streamwise) and
$\lambda_\theta = 2\pi/m$ in azimuth (spanwise).
Harmonic forcing is also considered in the following discussion.  
For a forcing of the form 
$\vec{f}(t) = \vec{\tilde{f}}\, {\mathrm e}^{{\mathrm i}\omega t}$, a harmonic response 
$\vec{u}(t) = \vec{\tilde{u}}\, {\mathrm e}^{{\mathrm i}\omega t}$ is expected.
The optimal response is then
\begin{equation}
    R(\alpha,m;\omega) = \max_{\vec{\tilde{f}}} \,
   \frac{||\vec{\tilde{u}}||}{||\vec{\tilde{f}}||} \, 
 \quad \mbox{ and } \quad
   R_{\mathrm{max}} = \max_\omega {R(\omega)} .
\end{equation}
The eigenfunction calculations were performed with up to $N=200$ points on $r\in[0,1]$
using a Chebyshev collocation method.
The number of (non-infinite) eigenvalues plus vectors obtained, 
being the number of degrees of freedom, is $2\,N-3$.  
Of these, 95\% were kept for the optimal growth analysis.  
At the highest $Re=10^6$, for which $Re_\tau=19200$, the power spectral drop-off of 
the optimal mode was of $8$ orders of magnitude 
(for both `inner' and `outer' modes, see below).

\section{Results}

The structure of optimal modes in strongly sheared flows is well known to 
reflect the lift-up mechanism, whereby rolls aligned with the flow
raise slow fluid from the boundary into the faster flow, leading to 
extended streaks.  While shearing is certainly present in the 
turbulent flow, it is subject to a significantly larger effective
viscosity, $\nu_T(r)$.

\begin{figure}[!b]
   \begin{center}
      \psfrag{Rem}{\small $Re=5300$}
      \psfrag{Ret}{\small $Re_\tau=187$}
      \psfrag{Gmax}{\small $G_{\mathrm{max}}$}
      \psfrag{m}{\small $m$}
      \psfrag{a=0}{\small $\alpha=0$}
      \epsfig{figure=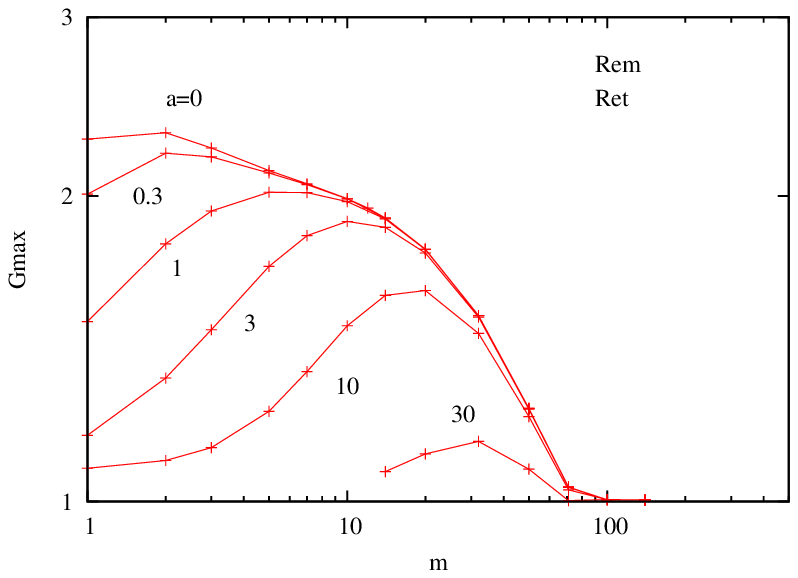, angle=0, scale=0.95}
      \psfrag{Rem}{\small $Re=10^6$}
      \psfrag{Ret}{\small $Re_\tau=19200$}
      \psfrag{Gmax}{\small $G_{\mathrm{max}}$}
      \psfrag{m}{\small $m$}
      \psfrag{a=0}{\small $\alpha=0$}
      \epsfig{figure=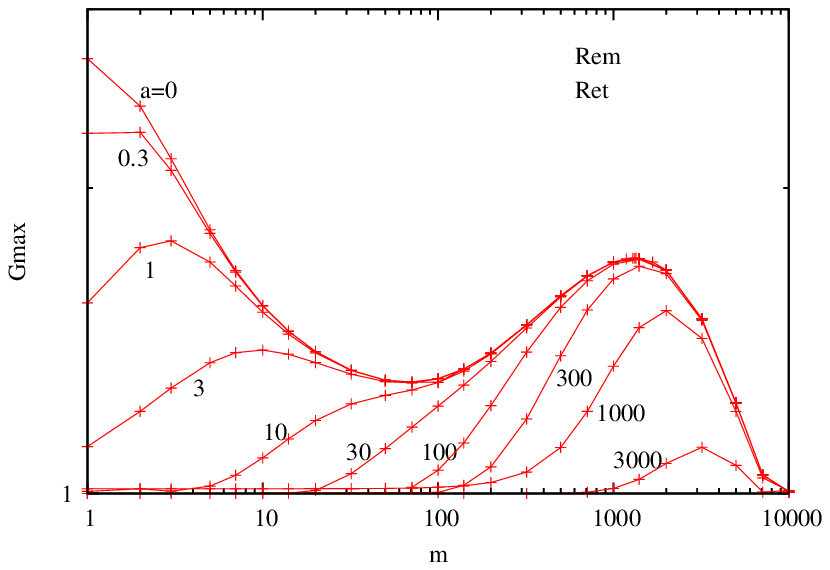, angle=0, scale=0.95}
   \end{center}
   \caption{\label{fig:5300Gmax}
      {\it Left}: Optimal growth at moderate Reynolds numbers.  While the 
      peak is for $m=1$, it is clear that for larger $\alpha$ 
      only modes of larger $m$ persist.
      {\it Right}: At large $Re$ a distinct peak appears at large $m$.
   }
\end{figure}

Optimal growths relative to the turbulent profile 
(figure \ref{fig:5300Gmax}) are significantly smaller than for the 
laminar case, although the greatest growth still occurs for the largest 
mode $m=1$.  For the turbulent case, however, $\nu_T$ reduces to the 
laminar value as one approaches the wall, and
growth of similar magnitude is also possible in the boundary region,
provided that the length scale is not so small that diffusion
again dominates.  

\begin{figure}
   \begin{center}
      \psfrag{2380}{\tiny $2380$}
      \psfrag{19200}{\tiny $19200$}
      \psfrag{Ret}{\tiny $Re_\tau=317$}
      \psfrag{Gmax}{\small $G_{\mathrm{max}}$}
      \psfrag{l+}{\small $\lambda_\theta^+$}
      \epsfig{figure=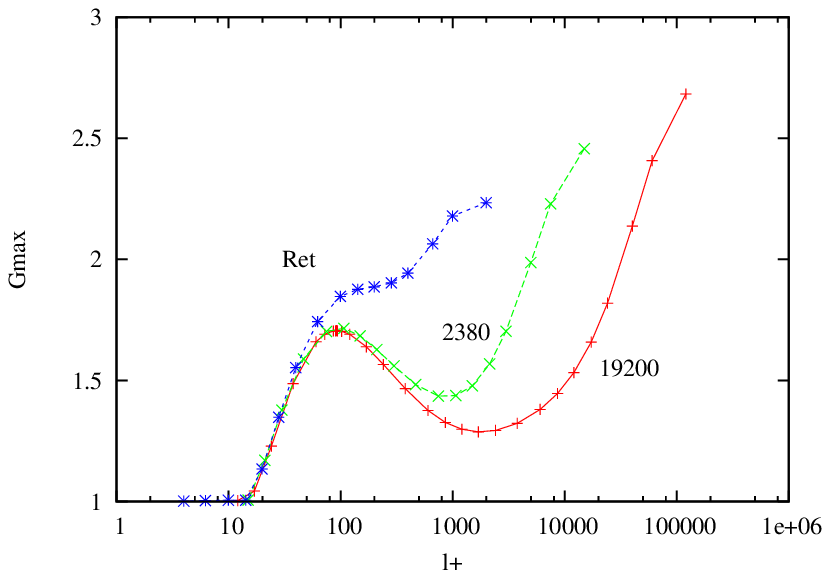, angle=0, scale=0.95}
      \psfrag{2380}{\tiny $2380$}
      \psfrag{19200}{\tiny $19200$}
      \psfrag{Ret}{\tiny $Re_\tau=317$}
      \psfrag{Tmax+}{\small $T^+_{\mathrm{max}}$}
      \psfrag{l+}{\small $\lambda_\theta^+$}
      \psfrag{l+/10}{\tiny\,\, $T^+=\lambda_\theta^+/10$}
      \epsfig{figure=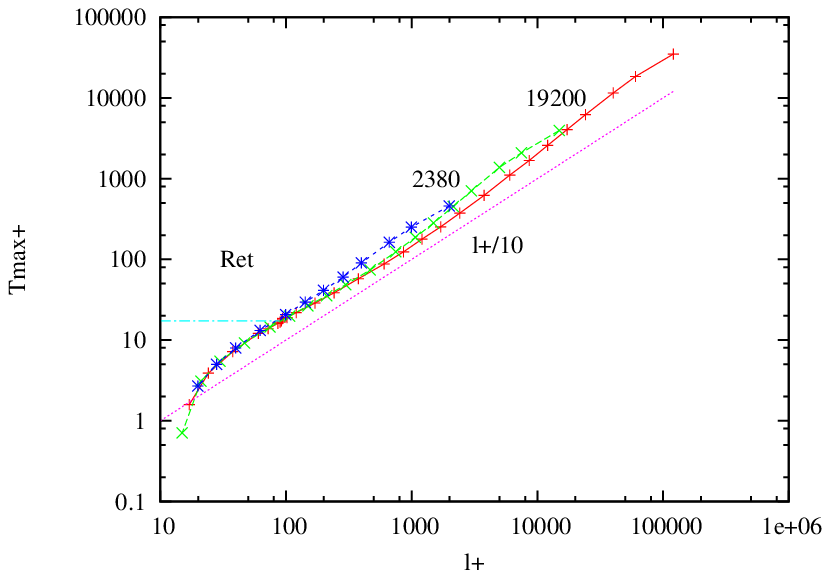, angle=0, scale=0.95}
   \end{center}
   \caption{\label{fig:l_inner}
      {\it Left}: An inner peak occurs for spanwise wavelength $\lambda_\theta^+=92$.
      {\it Right}: The time $T^+$ s.t. $G(T)=G_{\mathrm{max}}$.
   }
\end{figure}

This secondary peak, for which rolls are located close to the wall,
is shown in figure \ref{fig:l_inner} to scale in `inner' units, 
where in the absence of any other length scale, 
all units are inferred from the 
stress at the wall and the kinematic viscosity,
$r=r^+\nu/u_\tau$, $t=t^+\nu/u_\tau^2$.
It is clear from the figure that such scaling
is appropriate at high $Re_\tau$, where the $G_{\mathrm{max}}$ 
collapse with a peak for $\lambda^+_\theta=2\pi\,Re_\tau/m$ at $92$.

This peak in spanwise wavelength compares very favourably with experimental
results.  Using apparatus flush with the floor of the Utah desert \cite{klewicki95}
using smoke visualisation, a peak at was found at $\lambda^+=93$,
for a $Re_\tau$ as large as $1.5\times 10^6$.  
In \cite{smith83} similar values ($\pm10$) 
were observed for $Re_\tau$ in the range $1000$ to $5000$.  
Our predicted value $\lambda^+=92$ is in excellent agreement.
In the present study, this
secondary inner peak is observed to be distinct from the large-scale peak ($m=1$)
for $Re_\tau$ of order $1000$, and $\lambda_\theta^+$ being immediately
close to the above value from the outset.  Indeed, at lower $Re_\tau$, 
far before the peak becomes distinct for the $\alpha=0$ case, 
it is clear that modes of finite 
streamwise extent select similar $\lambda_\theta^+$.

\begin{figure}
   \begin{center}
      \epsfig{figure=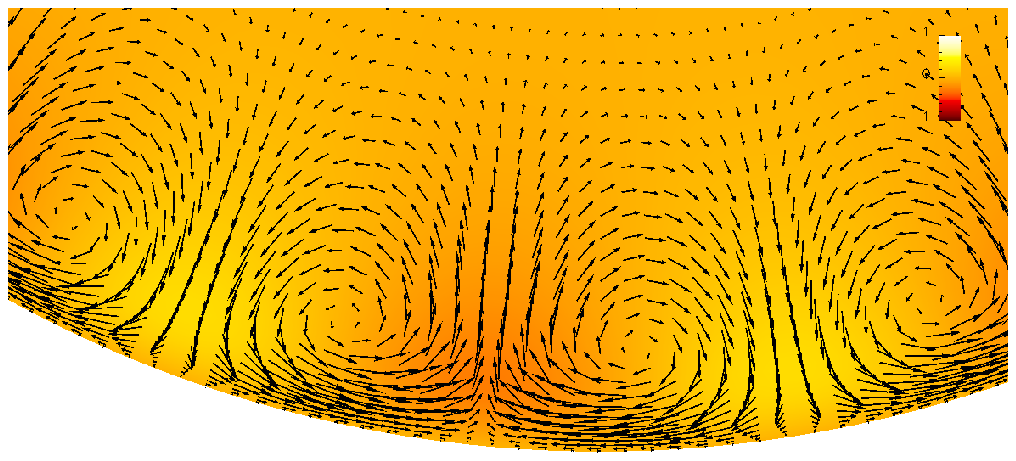, angle=0, width=60mm}
      \hspace{10mm}
      \epsfig{figure=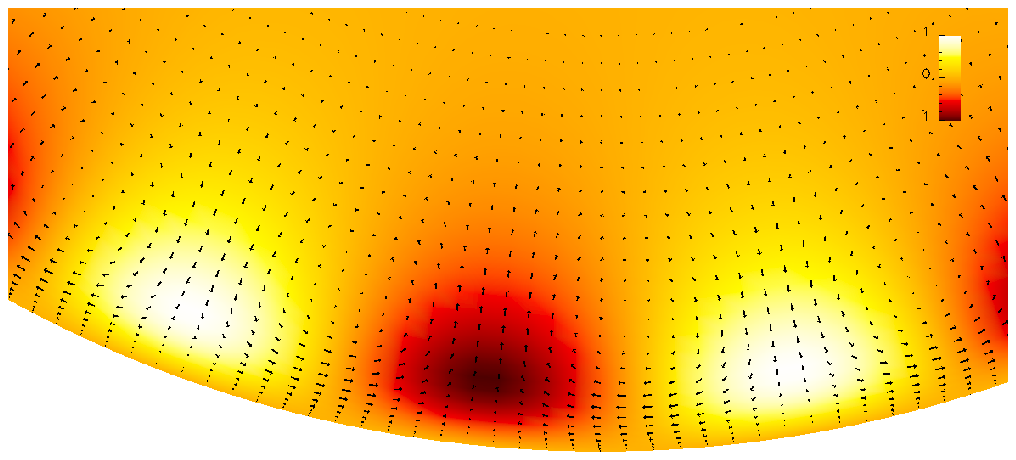, angle=0, width=60mm}
   \end{center}
   \caption{\label{fig:ModRes}
      Structure of the optimal inner-mode ({\it left}) and its 
      response ({\it right}).  
      Each is normalised by the maximum of $|\vec{u}|$.  
      The initial condition consists of rolls, and the response
      is dominated by streaks.
   }
\end{figure}

In figure \ref{fig:ModRes} the optimal initial condition and response
are plotted for the case $Re=5300$ ($Re_\tau\approx 180$) for $m=12$
($\lambda_\theta^+\approx 98$).
The rolls are typical of those of the wall mode 
at much greater $Re$.

\begin{figure}
   \begin{center}
      \psfrag{Rw}{\small $R(\omega)$}
      \psfrag{Rw+}{\small $R(\omega^+)$}
      \psfrag{w,w+}{\small $\omega,\,\,\omega^+$}
      \psfrag{m=1}{\small $\lambda_\theta=2\pi$}
      \psfrag{m=1314}{\small $\lambda_\theta^+=92$}
      \epsfig{figure=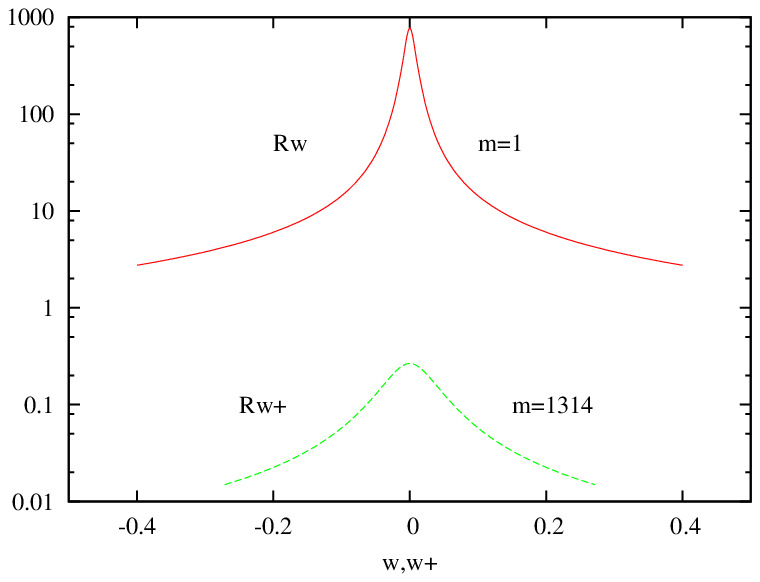, angle=0, scale=0.95}
      \psfrag{m}{\small $m$}
      \psfrag{Rmax}{\small $R_{\mathrm{max}}$}
      \psfrag{Gmax}{\small $G_{\mathrm{max}}$}
      \epsfig{figure=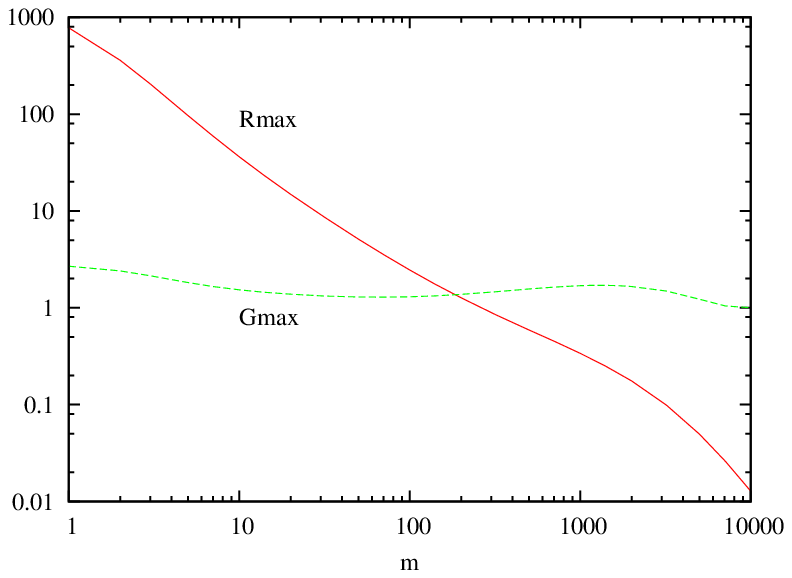, angle=0, scale=0.95}
   \end{center}
   \caption{\label{fig:Rwalp0}
      Response to harmonic forcing;  $Re_\tau=19200$.  
      For $\alpha=0$, steady 
      forcing is largest ($\omega=0$).  The outer mode ($m=1$) 
      is many orders more responsive for all moderate $\omega$
      than the inner mode ($m=1314$).
   }
\end{figure}

The growth of an initial condition is all very well, but how
easy is it to invoke such an initial condition?  The response 
to streamwise independent forcing is plotted over a range of 
harmonic frequencies, $\omega$, in figure \ref{fig:Rwalp0}.
The steady forcing, $\omega=0$, is most effective at generating a
response for this case.  While this is not true for non-zero $\alpha$,
the peak response remained smaller than that for the axially independent
case in all our calcuations.
The code described in \cite{willis09} has been modified
to include radially dependent viscosity.  The extreme difference 
in response to forcing different $m$ modes
was verified by timestepping relative to the 
prescribed mean profile.
The structure of the optimal force $\vec{\tilde{f}}$, however, is 
essentially identical to that of the optimal growth initial 
condition $\vec{u}_0$ (see figure \ref{fig:ModRes}); similarly 
for their respective responses.

\begin{figure}
   \begin{center}
      \psfrag{t+}{\small $t^+$}
      \psfrag{uf/f}{\small $||\vec{u}_f(t)|| \, / \, ||\vec{\tilde{f}}||$}
      \epsfig{figure=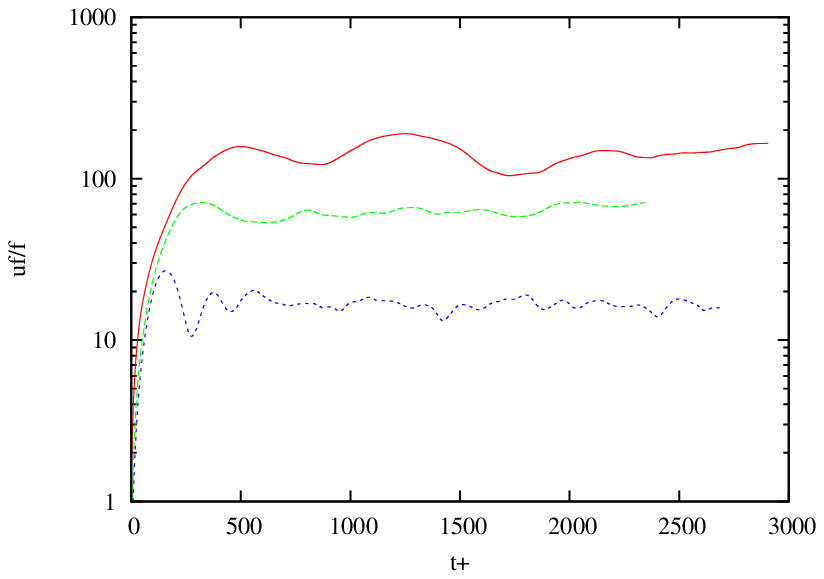, angle=0, scale=0.95}
      \psfrag{t+}{\small $t^+$}
      \psfrag{Cf}{\small $C_f$}
      \epsfig{figure=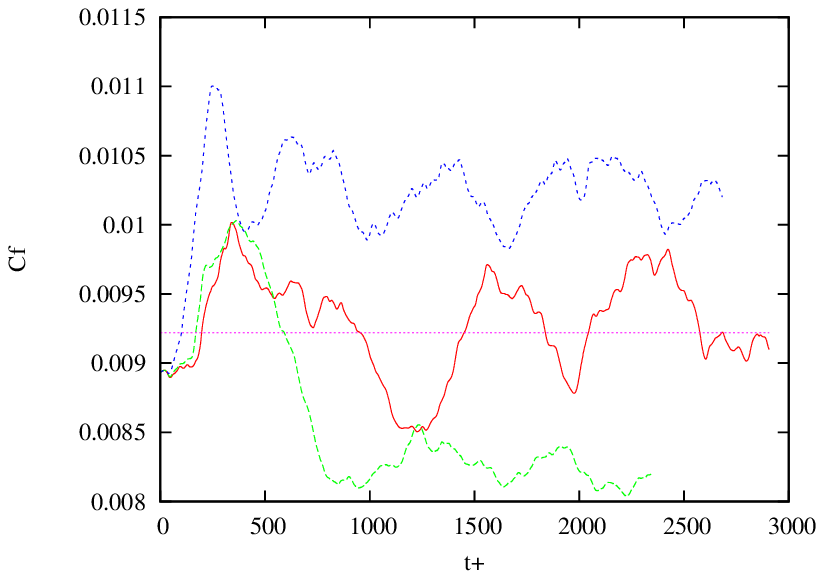, angle=0, scale=0.95}
   \end{center}
   \caption{\label{fig:5300Rspnc}
      Response of a turbulent flow to forcing;  
      $Re_\tau\approx 180$, $\alpha=0, m=1$ and 
      $||\vec{\tilde{f}}||^2=10^{-6},10^{-5},10^{-4}$.  
      Here $||\cdot||^2 = \frac{1}{2}\int |\cdot|^2 {\mathrm d}V$.
      The expected
      $R_{\mathrm{max}}$ is 280.  A similar order value is achieved for
      $||\vec{\tilde{f}}||^2=10^{-6}$ and decreases as the force increases.
      The amplitude of the induced flow is already as large as 
      5 to 10\% of the mean flow for all cases.  The forcing has a 
      significant affect on the skin-friction coefficient 
      $C_f=2(u_\tau/U_b)^2$, the horizontal line being the unforced average.
   }
\end{figure}

Unlike the transient growth, steady forcing provides a 
convenient platform for testing the model, 
whereby statistical measures averaged over much longer times may 
be accumulated.  
Consider the timestepped velocity field, $\vec{u}(t)$, obtained 
from direct numerical simulation of the original Navier--Stokes 
equations subject to a forcing $\vec{\tilde{f}}$. 
Then
$\vec{u}_f(t)=(\vec{u}(t)\cdot\vec{\hat{\tilde{u}}})\,\vec{\hat{\tilde{u}}}$
is the component of $\vec{u}(t)$ in the direction of the the expected 
normalised response field $\vec{\hat{\tilde{u}}}$.
The observed quantity $||\vec{u}_f(t)||/||\vec{\tilde{f}}||$ may 
be compared with the expected response $R_{\mathrm{max}}$.
Figure \ref{fig:5300Rspnc} shows the response of the turbulent flow
$Re=5300$, computed in a domain of length $10$ radii ($\lambda_z^+\approx1800$)
at a resolution $(60,\pm64,\pm64)$ before dealiasing.
The calculations clearly demonstrate that large amplification may occur,
as predicted by the model.  Note also that for intermediate forcing, drag reduction
is possible.   

\section{Conclusions}

The model has been shown to successfully predict 
large growth, to good quantitive agreement, despite 
the assumptions of linearity upon a turbulent state and
the isotropic eddy viscosity assumption.  It shows the huge
response to forcing of large scale modes, and accurately
predicts the peak spanwise wavelength, $\lambda_\theta^+=92$, of
the near-wall modes.

Analogue studies for optimal growth in 
turbulent channels \cite{delalamo06,pujals09}, 
the boundary layer \cite{cossu09}
and the Couette flow \cite{hwang09}, 
all show close spanwise wavelengths for the wall mode.
It is surprising, however, that the optimal growth 
calculations suggest small growth of the wall modes, 
despite clear observational evidence that such structures exist.

The results from harmonic forcing, on the other hand, 
suggest that the large scale modes are easily excited.
Such relative difficulty of forcing motion on small scales 
has consequences for control of turbulence.  It clearly
requires considerable effort to locally manipulate structures 
in the neighbourhood of the wall.  It is possible that it is more 
straight forward to control indirectly via forcing of larger, or
possibly even very large scales, taking advantage of the 
much greater linear response.  Indeed, such a possiblity has 
been shown in principle for channel flow \cite{schoppa98}, and is
verified here for a significantly larger ratio of `large' 
(imposed roll) to `small' (streak spacing) scales.
The nature of suppression by a large-scale motion deserves
further investigation.


\bibliographystyle{unsrt-cfm}
\bibliography{trans}
\end{document}